 \documentclass[prb,twocolumn,showpacs,showkeys]{revtex4}

\usepackage{graphicx}
\usepackage{amssymb,amsmath}
\usepackage{dcolumn}
\usepackage{bm}
\usepackage{color}

\begin{document}

\title{Magnetic ordering in the static intermediate-valent cerium 
       compound $ {\bf Ce_2RuZn_4} $}

\author{Volker Eyert}
\email[Corresponding author. \\
       Email-address:]{eyert@physik.uni-augsburg.de} 
\affiliation{Center for Electronic Correlations and Magnetism,
             Institut f\"ur Physik, Universit\"at Augsburg,
             D-86135 Augsburg, Germany}
\author{Ernst-Wilhelm Scheidt}
\affiliation{Chemische Physik und Materialwissenschaften, Institut f\"ur 
             Physik, Universit\"at Augsburg, D-86135 Augsburg, Germany}
\author{Wolfgang Scherer}
\affiliation{Chemische Physik und Materialwissenschaften, Institut f\"ur 
             Physik, Universit\"at Augsburg, D-86135 Augsburg, Germany}
\author{Wilfried Hermes}
\affiliation{Institut f\"ur Anorganische und Analytische Chemie, 
             Westf\"alische Wilhelms-Universit\"at M\"unster, 
             D-48149 M\"unster, Germany}
\author{Rainer P\"ottgen}
\affiliation{Institut f\"ur Anorganische und Analytische Chemie, 
             Westf\"alische Wilhelms-Universit\"at M\"unster, 
             D-48149 M\"unster, Germany}

\date{\today}

\begin{abstract}
The low-temperature behavior of $ {\rm Ce_2RuZn_4} $ has been investigated. 
Specific heat and magnetic susceptibility data reveal an antiferromagnetic 
transition at a N\'eel temperature of 2\,K. $ {\rm Ce_2RuZn_4} $ is a 
static intermediate-valent compound with two crystallographically 
independent cerium atoms. The magnetic data clearly show that only one 
cerium site is magnetic ($ {\rm Ce^{3+}} $), while the second one 
carries no magnetic moment. The experimental 
data are interpreted with the help of first principles electronic structure 
calculations using density functional theory and the augmented spherical 
wave method. The calculations reveal the occurrence of two different 
cerium sites, which are characterized by strongly localized magnetic 
moments and strong Ce-Ru bonding, respectively. 
\end{abstract}



\maketitle

\section{Introduction}
\label{intro}

Ternary intermetallic compounds in the system Ce-Ru-X, where X is an
element of the 3rd, 4th, or 5th main group, have been intensively 
studied in recent years. This is due to the fact that these materials 
exhibit peculiar structural and physical properties, where the 
transition metal ruthenium seems to play a key role. In comparison 
to related compounds with other transition metals, ruthenium often 
shows extremely short Ce-Ru distances, even much shorter than the sum 
of the covalent radii of 289\,pm. \cite{emsley99} Already the binary 
Laves phase $ {\rm CeRu_2} $, \cite{compton59} which is a superconductor 
below 6.2\,K, \cite{wilhelm70} is extraordinary. In this compound, 
the cerium atoms are in an intermediate valent state and the $ a $ 
lattice parameter is much smaller than that of $ {\rm NdRu_2} $. When 
it comes to the ternary compounds, these effects are even more pronounced. 
The recently reported indium compounds $ {\rm CeRu_{0.88}In_2} $, 
$ {\rm Ce_{16}Ru_8In_{37}} $, $ {\rm Ce_3Ru_2In_2} $, and 
$ {\rm Ce_3Ru_2In_3} $ exhibit extremely short Ce-Ru distances, 
\cite{kurenbaeva07,murashova07,tursina07,murashova08} which are directly 
associated with a strong tendency of some of the cerium sites towards 
tetravalency.

The stannide $ {\rm CeRuSn} $ also shows this peculiar behavior. 
\cite{riecken07,matar07} The room-temperature modification adopts a 
superstructure of the $ {\rm CeCoAl} $ type and the two crystallographically 
independent cerium sites are ordered on distinct positions. In particular, 
the intermediate valent Ce1 atoms have very short Ce1-Ru distances of 
only 233-246\,pm. Consequently, our electronic structure calculations 
revealed strong Ce-Ru bonding. 

Another example for trivalent/intermediate valent cerium ordering is 
the structure of $ {\rm Ce_2RuZn_4} $. \cite{mishra08} This structure 
type crystallizes in the space group $ P4/nmm $ with two formula units 
per cell and two twofold cerium sites with distinctly different 
coordination. Preliminary magnetic susceptibility investigations 
underlined the idea of trivalent/intermediate valent cerium ordering. 
Extending our systematic studies of such ruthenium based intermetallic 
compounds we here report on a comprehensive investigation of the 
low-temperature properties of $ {\rm Ce_2RuZn_4} $. In particular, 
our work includes X-ray diffraction, specific heat and magnetic 
susceptibility measurements as well as electronic structure calculations.

\section{Experimental}
\label{exp}

\subsection{Synthesis}
\label{expsyn}

Starting chemicals for the preparation of $ {\rm Ce_2RuZn_4} $ were 
a cerium ingot (Johnson Matthey, purity $ > $ 99.9 \%), ruthenium 
powder (Degussa-H\"uls, ca.\ 200 mesh, purity $ > $ 99.9 \%), and 
zinc granules (Merck, purity $ > $ 99.9 \%). Pieces of the cerium 
ingot were first arc-melted to a small button under argon. 
\cite{poettgen99} The argon was purified 
before with molecular sieve, silica gel and titanium sponge (900\,K). 
The elements were then mixed in the 2:1:4 atomic ratio (ruthenium 
powder was cold-pressed to a pellet of 6\,mm diameter) and arc-welded 
in a tantalum ampoule under an argon pressure of 700\,mbar. The ampoule 
was subsequently placed in a water-cooled sample chamber of a 
high-frequency furnace (H\"uttinger Elektronik, Freiburg, Typ TIG 5/300). 
\cite{kussmann98} The heat treatment was similar to our 
previous experiments: \cite{mishra08} three times for 1\,min at 
1360\,K, for 10\,min at 1360\,K, rapid cooling to 920\,K, and a final 
annealing at that temperature for another 2 hours followed by quenching.

\subsection{Powder X-ray Diffraction}
\label{expxrd}

The purity of the sample was checked through a Guinier powder pattern 
using Cu-$ {\rm K}_{\alpha1} $ radiation and $ \alpha $-quartz 
($ a $ = 491.30\,pm, $ c $ = 540.46\,pm) as an internal standard. The 
tetragonal lattice parameters were obtained from a least-squares 
refinement of the Guinier powder data. The refined lattice parameters 
($ a = 719.6(1) $\,pm, $ c = 520.0(1) $\,pm) are in good agreement 
with the previously reported data ($ a $ = 719.6(1)\,pm, 
$ c $ = 520.2(1)\,pm). \cite{yvon77} No impurity phases have been 
detected on the level of X-ray powder diffraction.

\subsection{Specific Heat and Magnetic Susceptibility}
\label{expppm}

The temperature-dependent heat capacity was investigated by means of 
a quasi-adiabatic step heating technique employing a Quantum Design 
PPMS system. The measurements were performed on a polycrystalline 
sample thermally connected by grease (Apiezon-N) at temperatures 
ranging from 1.8 up to 300\,K. For temperatures below 2\,K, down to 
about 80\,mK, specific heat data were also collected in a 
$ {\rm ^3He} $/$ {\rm ^4He} $-dilution-cryostat using a relaxation 
method. \cite{bachmann72} The uncertainty of the measurements reported 
in this paper is estimated to be below 3\%. The magnetic measurements 
in the temperature range 1.8\,K $ < T < $ 300\,K were done using a 
Quantum Design MPMS7 superconducting quantum interference device (SQUID) 
magnetometer.

\section{Theoretical Method}
\label{theo}

The calculations are based on density-functional theory (DFT) and
the local density approximation (LDA). They were performed using
the scalar-relativistic implementation of the augmented spherical
wave (ASW) method (see Refs.\ \onlinecite{aswrev,aswbook} and
references therein). In the ASW method, the wave function is
expanded in atom-centered augmented spherical waves, which are
Hankel functions and numerical solutions of Schr\"odinger's
equation, respectively, outside and inside the so-called
augmentation spheres. In order to optimize the basis set,
additional augmented spherical waves were placed at carefully
selected interstitial sites. The choice of these sites as well as
the augmentation radii were automatically determined using the
sphere-geometry optimization algorithm. \cite{sgo} Self
consistency was achieved by a highly efficient algorithm for
convergence acceleration. \cite{mixpap} The Brillouin zone
integrations were performed using the linear tetrahedron method
with up to 105 {\bf k}-points within the irreducible wedge.  
\cite{bloechl94,aswbook} 

In the present work, we used a new full potential version of the
ASW method, which was implemented only very recently. \cite{fpasw}
In this version, the electron density and related quantities are
given by a spherical harmonics expansion inside the muffin-tin
spheres. In the remaining interstitial region, a representation in
terms of atom-centered Hankel functions is used. \cite{msm88}
However, in contrast to previous related implementations, we here
get away without needing a so-called multiple-$ \kappa $ basis
set, which fact allows for a very high computational speed of 
the resulting scheme.

\section{Results and Discussion}
\label{results}

\subsection{Crystal Chemistry}
\label{cryschem}

\begin{figure}[htb]
\centering
\includegraphics[width=1.0\columnwidth,clip]{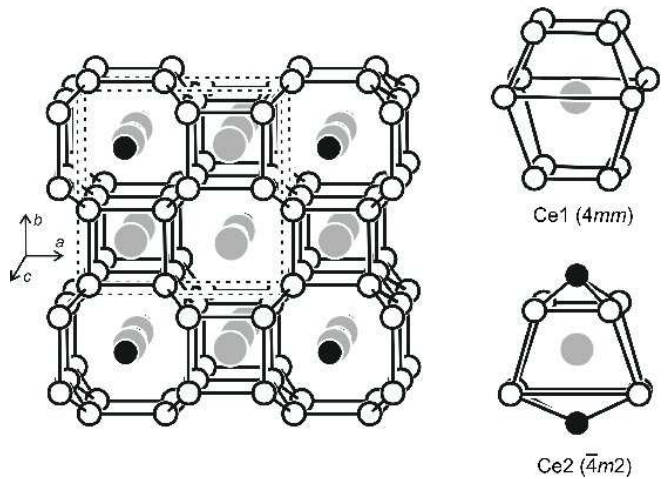}
\caption{The crystal structure of $ {\rm Ce_2RuZn_4} $: 
         (left-hand side) view of the structure approximately along the 
         $ c $ axis and (right-hand side) coordination polyhedra and 
         site symmetries of the two crystallographically independent 
         cerium sites. The cerium, ruthenium, and zinc atoms are drawn 
         as medium grey, filled, and open circles, respectively.} 
\label{fig:struct}
\end{figure}

$ {\rm Ce_2RuZn_4} $ crystallizes with a peculiar structure type 
with space group $ P4/nmm $. \cite{mishra08} Since the crystal chemistry 
of this zinc-rich intermetallic compound has been described in detail 
in our previous work, \cite{mishra08} here we focus only on the cerium 
coordination, which is responsible for the distinctly different magnetic 
behaviour of the two cerium sites. A view of the $ {\rm Ce_2RuZn_4} $ 
structure approximately along the $ c $ axis is presented in Fig.\ 
\ref{fig:struct}.  
%
The zinc atoms build up a complex three-dimensional network, which 
leaves two different kinds of channels parallel to the $ c $ axis. 
They are filled by the cerium and ruthenium atoms. As emphasized in 
the right-hand part of Fig.\ \ref{fig:struct}, the trivalent Ce1 
atoms have 12 nearest neighbours of only the zinc type. In contrast, 
the intermediate-valent (nearly tetravalent) Ce2 atoms have a smaller 
coordination number of 10 atoms with two ruthenium and eight zinc sites. 
However, the most striking structural feature are the short Ce2-Ru 
distances of 260\,pm, which are even shorter than the sum of the covalent 
radii of 289\,pm. \cite{emsley99} This seems to be a characteristic of 
ruthenium and leads to a destabilization of the cerium valence. The 
peculiar electronic properties, which arise from this unusual crystal 
chemistry, are addressed below.

\subsection{Specific Heat}
\label{specheat}

\begin{figure}[htb]
\centering
\includegraphics[width=1.0\columnwidth,clip]{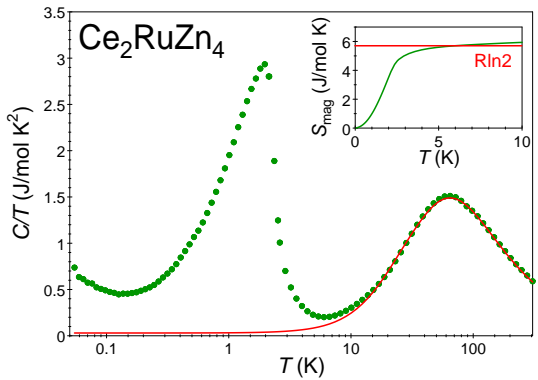}
\caption{(Color online) Specific heat divided by temperature vs.\ 
          $ \log T $ of $ {\rm Ce_2RuZn_4} $ in the temperature range 
          between 50\,mK and 300\,K. The slight increase of $ C/T $ 
          below 1\,K is mainly due to the nuclear quadrupole moments 
          displayed by the $ {\rm ^{101}Ru} $ and $ {\rm ^{67}Zn} $ 
          isotopes; the solid red line represents the lattice and 
          electronic contributions. The inset shows the magnetic 
          entropy. The red line displays $ S_{mag} = R \ln 2 $, which 
          is the limit for a two level system.} 
\label{fig:cpexp}
\end{figure}

In Fig.\ \ref{fig:cpexp},  
the molar specific heat capacity divided by temperature $ C/T $ of 
$ {\rm Ce_2RuZn_4} $ is displayed in a semi-logarithmic plot for 
temperatures between 0.08 and 300\,K. A clear peak at 2\,K indicates 
a pronounced magnetic contribution to the heat capacity. In order to 
extract the magnetic contribution of the specific heat we determined 
the lattice contributions theoretically using a single Debye term and 
two Einstein modes. In this calculation, we fixed the number of internal 
degrees of freedom to 21, according to the seven atoms in the unit 
cell. Using a weight distribution of 1:2:4, we obtained a Debye 
temperature of $ \Theta_{\rm D} = 96 $\,K, and two Einstein modes of 
$ \Theta_{\rm E1} = 129 $\,K and $ \Theta_{\rm E2} = 238 $\,K. The solid 
red line in Fig.\ \ref{fig:cpexp} shows the corresponding result 
taking into account 
an electronic contribution of $ \gamma = 30 $\,$ {\rm {mJ/molK^2}} $. 
Below 1\,K, the specific heat of $ {\rm Ce_2RuZn_4} $ increases 
slightly giving rise to a nuclear electronic Schottky contribution. 
We have calculated the hyperfine contributions to the heat capacity 
originating in zero magnetic field mainly from the quadruple moments 
of $ {\rm ^{101}Ru} $ and $ {\rm ^{67}Zn} $ leading to an average 
local internal electrical field gradient of 
$ V_{zz} = (5 \pm 0.8) \times 10^{22} $\, $ {\rm V/m^2} $.

Subtracting the calculated electronic, lattice and nuclear parts from 
the total specific heat data results in the magnetic contribution 
of specific heat $ C_{mag} $. Subsequent integration of $ C_{mag}/T $ 
leads to the magnetic entropy according to:
\begin{displaymath} 
S_{mag} = \int \frac{C_{mag}}{T} dT = R \ln (2S+1) 
                        \;.
\end{displaymath} 
Here, $ S_{mag} $ and $ S $ refer to the magnetic entropy and the spin, 
respectively. The corresponding graph is shown in the inset of Fig.\
\ref{fig:cpexp}. The resulting magnetic entropy of $ R \ln 2 $ is due 
to a two level $ S = 1/2 $ state.

\subsection{Magnetic Susceptibility}
\label{suscept}

\begin{figure}[htb]
\centering
\includegraphics[width=1.0\columnwidth,clip]{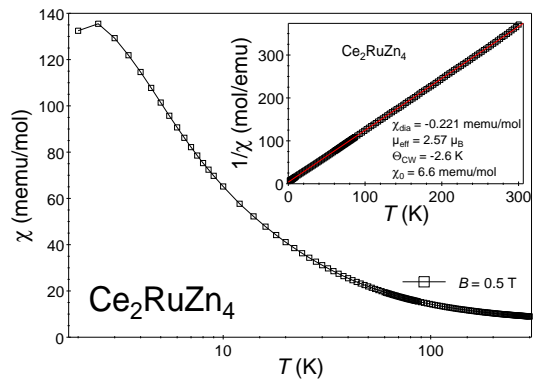}
\caption{(Color online) The magnetic susceptibility $ \chi $ vs.\ 
          $ \log T $ of $ {\rm Ce_2RuZn_4} $. At low temperature the 
          beginning of an antiferromagnetic transition is observed. 
          The inset exhibits the inverse susceptibility $ 1/\chi $ 
          vs.\ $ T $. The solid line is a least squares fit according 
          to the modified Curie-Weiss-law.} 
\label{fig:chiexp}
\end{figure}

The magnetic configuration of the Ce ion can be proven by susceptibility 
measurements $ \chi (T) $. Our susceptibility data collected at a field 
of 0.5\,T are shown in Fig.\ \ref{fig:chiexp}. 
Here the molar core diamagnetism of $ \chi_{dia} = - 0.221 $\,memu/mol 
is substracted from the data. $ \chi_{dia} $ is calculated using the 
expression 
$ \chi_{dia} = - 0.79 \times \sum_i Z_i \times 10^{-6} (r/a_B)^2 $\,emu/mol, 
where $ \sum_i Z_i $ is the sum of all core electrons per formula unit, 
$ (r/a_B) $ is roughly estimated to be unity, and $ a_B $ is the 
Bohr radius. In the inset of Fig.\ \ref{fig:chiexp}, a linear 
temperature dependence of $ \chi^{-1} (T) $ is observed above 50\,K. 
A fit according to the modified Curie–Weiss-law 
\begin{displaymath} 
\chi = \frac{C}{ T - \Theta_{\rm CW} } + \chi_0 
\end{displaymath} 
yields a temperature independent susceptibility $ \chi_0 = 6.6 $\,memu/mol, 
an effective moment $ \mu_{eff} = 2.57 \mu_B $ and the paramagnetic 
Curie-Weiss temperature $ \Theta_{\rm CW} = - 2.6 $\,K. The effective 
magnetic moment is close to the theoretical value associated with the 
$ {\rm Ce^{3+}} $ state, i.e.\ $ 2.54 \mu_B $, but not to the calculated 
value of $ \mu_{eff} = 3.59 \mu_B $ as it is expected for the two 
$ {\rm Ce^{3+}} $ ions in the unit cell. These data clearly indicate 
that in $ {\rm Ce_2RuZn_4} $ the Ce ions occur in two different 
valence states $ {\rm Ce^{3+}} $ and $ {\rm Ce^{4+}} $. A slight 
curvature of the magnetic susceptibility observed below 3\,K is 
attributed to the beginning of an antiferromagnetic transition with 
a N\'eel-temperature of $ T_{\rm N} = 2 $\,K as already derived from 
the specific heat data.

\subsection{Electronic Structure}
\label{chembond}

The investigation of the electronic structure was performed in two 
steps. While in a first set of calculations, spin-degeneracy was 
enforced, we considered spin-polarization only at a later stage. 
This allowed to investigate the influence of the magnetic order on 
the electronic states and the total energy in more detail. 

\begin{figure}[htb]
\centering
\includegraphics[width=1.0\columnwidth,clip]{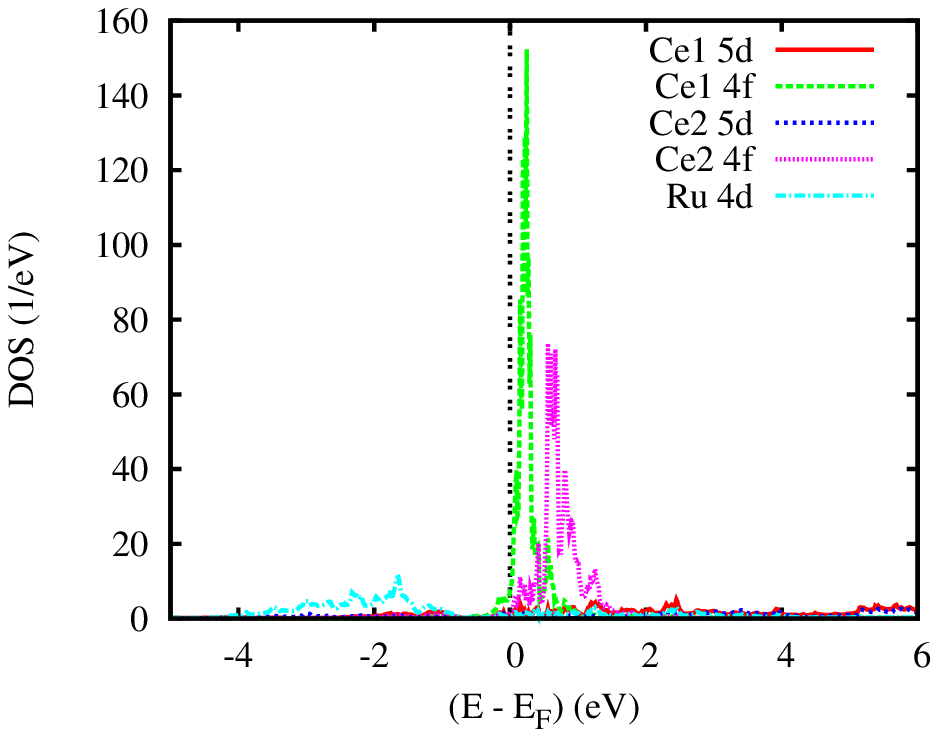}
\caption{(Color online) Partial densities of states of non-magnetic 
         $ {\rm Ce_2RuZn_4} $.} 
\label{fig:dosnm1}
\end{figure}

\begin{figure}[htb]
\centering
\includegraphics[width=1.0\columnwidth,clip]{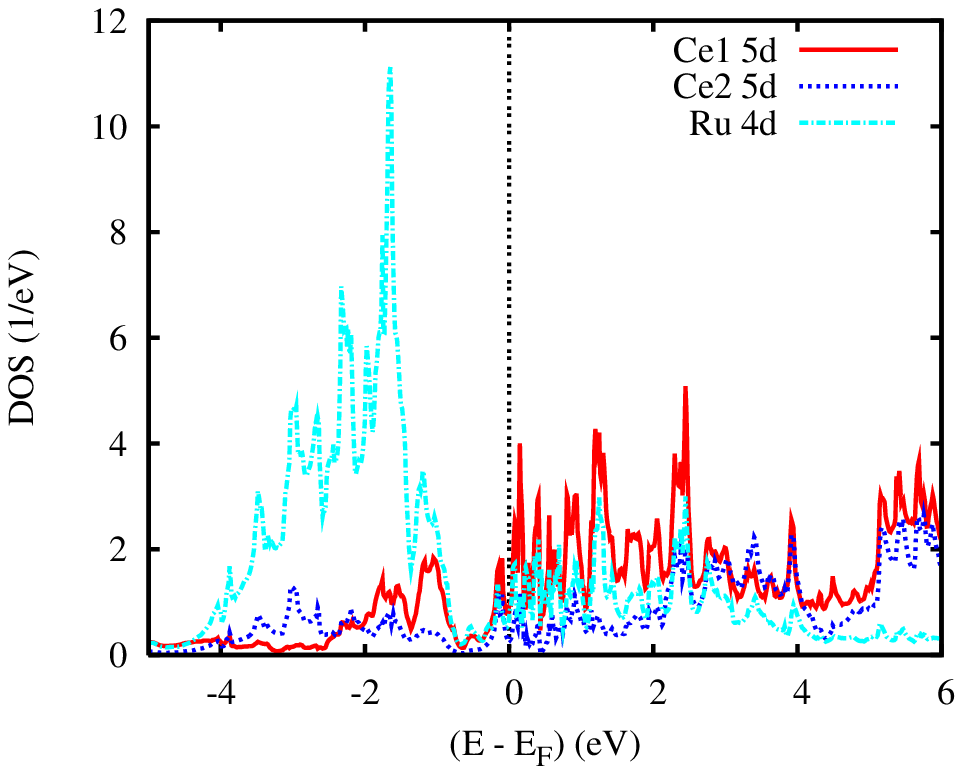}
\caption{(Color online) Partial densities of states of non-magnetic 
         $ {\rm Ce_2RuZn_4} $.} 
\label{fig:dosnm2}
\end{figure}

The partial densities of states (DOS) resulting from the first step 
are displayed in Figs.\ \ref{fig:dosnm1} 
and \ref{fig:dosnm2}. 
According to Fig.\ \ref{fig:dosnm1}, the densities of states is 
dominated by the cerium $ 4f $ states. The different behaviour 
of the two cerium sites is clearly visible. While the $ 4f $ 
states of the Ce1 atoms, which are at the centers of the Zn cages, 
form a very sharp peak right above the Fermi energy, the linearly 
coordinated Ce2 $ 4f $ states give rise to rather broad bands of 
about 1\,eV width due to their overlap with the Ru $ 5d $ states. 

The partial DOS due to the cerium and ruthenium $ d $ states are 
displayed in Fig.\ \ref{fig:dosnm2}. As expected, the Ru $ 4d $  
states are located well below the Fermi energy. In contrast, the 
Ce $ 5d $ states lead to broad bands starting just slightly below 
$ E_{\rm F} $. Yet, hybridization of both types of bands leads to 
considerable contributions of the cerium and ruthenium states, 
respectively, below and above $ {\rm E_F} $. In particular, we 
observe a pronounced similarity of the Ru $ 4d $ partial DOS with 
that due to the Ce $ 5d $ states, which is a signature of strong 
$ d $-$ d $ overlap, which is even stronger for the Ce2 sites. 
In passing, we mention the Zn $ 3d $ states, which lead to a narrow 
band in the energy range from $ -8 $ to $ -7 $\,eV and are not 
included in the plot. 

\begin{figure}[htb]
\centering
\includegraphics[width=1.0\columnwidth,clip]{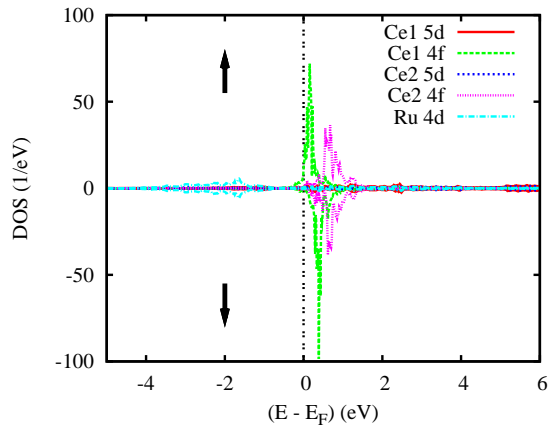}
\caption{(Color online) Partial densities of states of assumed 
         ferromagnetic $ {\rm Ce_2RuZn_4} $.} 
\label{fig:dosfe}
\end{figure}

For simplicity reasons, subsequent spin-polarized calculations were 
performed for an assumed ferromagnetic order. These calculations  
resulted in a lowering of the total energy by about 0.9\,mRyd and 
a magnetic moment of about 
$ {\rm 0.4 \mu_B} $ per formula unit, respectively. Interestingly, 
the latter is carried by the Ce1 atoms alone, whereas the Ce2 states 
are not polarized. This becomes obvious from Fig.\ \ref{fig:dosfe}, 
which displays the partial densities of states of assumed ferromagnetic 
$ {\rm Ce_2RuZn_4} $. Obviously, the finite spin-polarization at 
the Ce1 sites, which is mainly due to the $ 4f $ states, results 
from the near-$ E_{\rm F} $ position of the latter. In contrast, 
the Ce2 $ 4f $ states are not polarized due to their much lower 
occupation. 

\begin{figure}[htb]
\centering
\includegraphics[width=1.0\columnwidth,clip]{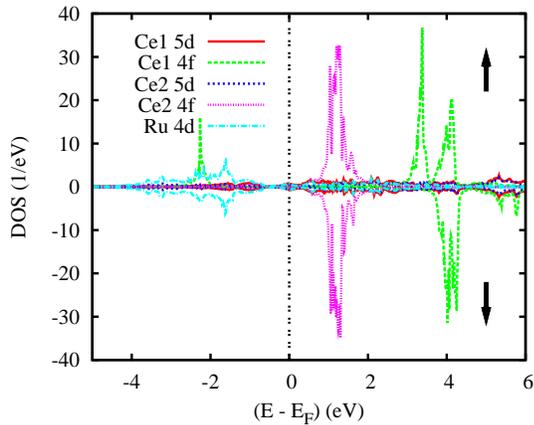}
\caption{(Color online) Partial densities of states of assumed 
         ferromagnetic $ {\rm Ce_2RuZn_4} $.} 
\label{fig:dosfeu}
\end{figure}

Motivated by the strongly localized nature of the Ce $ 4f $ states, 
additional LDA+U calculations were performed with $ U = 6.8 $\,eV 
and $ J = 0.68 $\,eV applied at both cerium sites. These calculations 
resulted in a magnetic moment of $ {\rm 1.0 \mu_B} $ at each Ce1 site 
as is expected for $ {\rm Ce^{3+}} $. In contrast, the Ce2 sites 
remain unpolarized. In addition, a complete restructuring of the 
electronic states as compared to the LDA calculations is observed 
in the partial densities of states shown in Fig.\ \ref{fig:dosfeu}. 
While the Ce2 $ 4f $ states remain centered at about 1\,eV, the 
Ce1 $ 4f $ states experience a considerable splitting of about 
6\,eV due to the strong electronic correlations. While the 
spin-minority states are essentially empty, spin-majority states 
display a finite DOS at about $ - 2 $\,eV, which leads to the 
magnetic moments at this site. In contrast, the $ d $ states of 
all species remain essentially unpolarized. 

\begin{figure}[htb]
\centering
\includegraphics[width=1.0\columnwidth,clip]{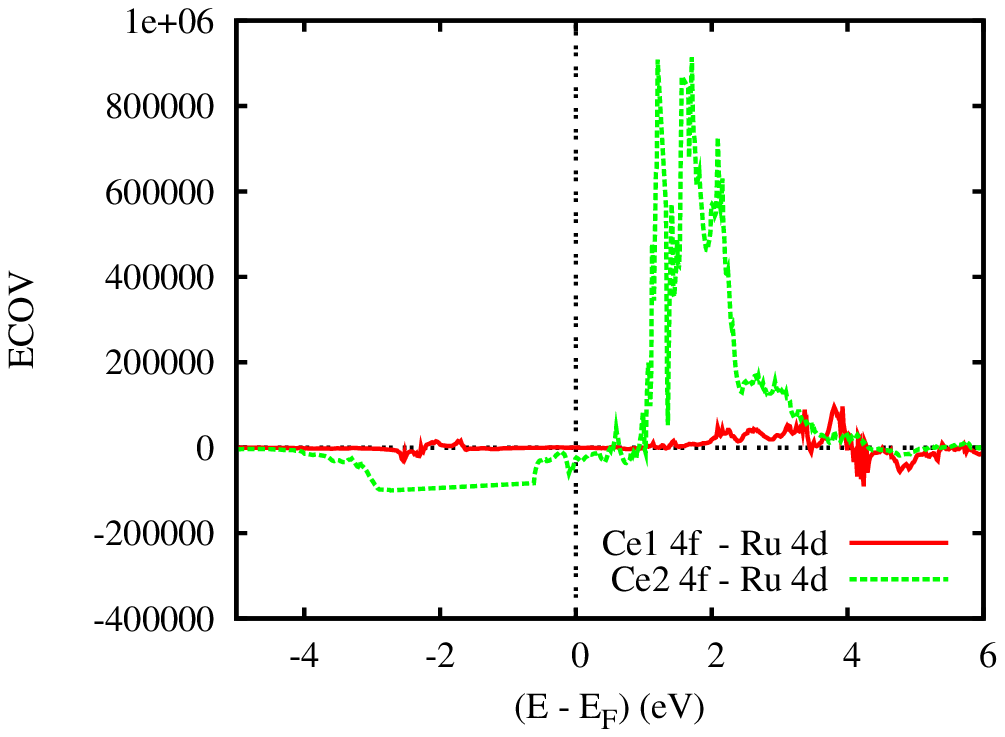}
\caption{(Color online) Partial covalence energies of assumed 
         ferromagnetic $ {\rm Ce_2RuZn_4} $.} 
\label{fig:coopfeuMfd}
\end{figure}

\begin{figure}[htb]
\centering
\includegraphics[width=1.0\columnwidth,clip]{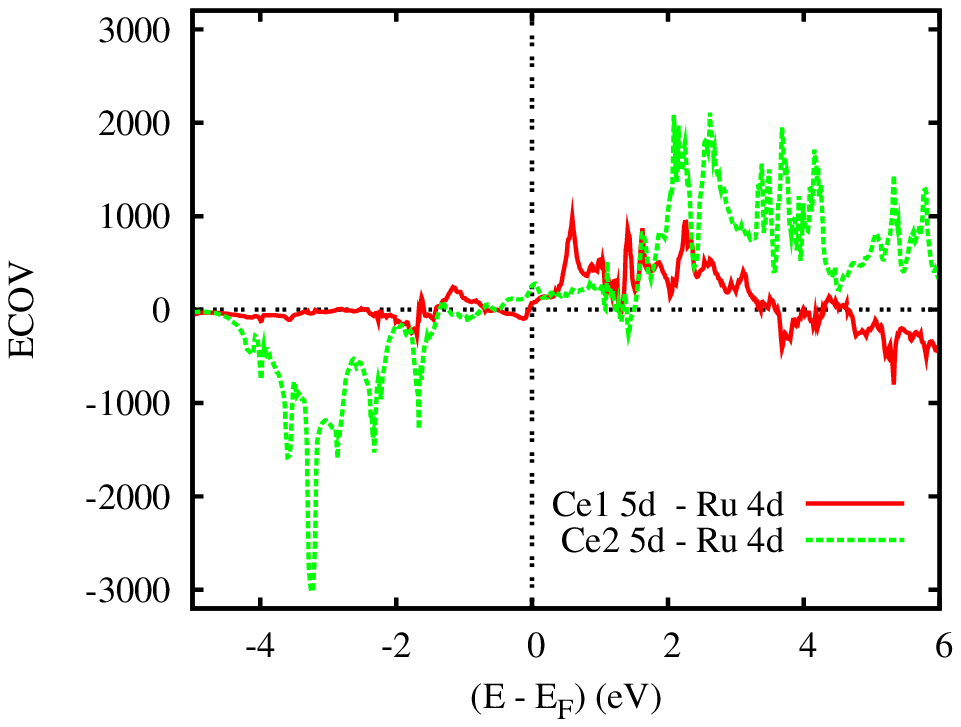}
\caption{(Color online) Partial covalence energies of assumed 
         ferromagnetic $ {\rm Ce_2RuZn_4} $.} 
\label{fig:coopfeuMdd}
\end{figure}

Finally, in order to investigate the differences in chemical bonding, 
we display in Figs.\ \ref{fig:coopfeuMfd} 
and \ref{fig:coopfeuMdd}
the partial covalence energies $ E_{cov} $ of assumed ferromagnetic 
$ {\rm Ce_2RuZn_4} $. In doing so, we concentrate on the bonding 
between the Ru $ 4d $ states with the $ 5d $ and the $ 4f $ states 
at the two different cerium sites. Note that negative and positive 
contributions to $ E_{cov} $ point to bonding and antibonding states, 
respectively. The results give a very clear and impressive indication 
of the strong bonding between the Ru atoms and the Ce2 sites; the 
$ E_{cov} $ curves are positive throughout up to $ E_{\rm F} $. In 
contrast, there is almost no bonding between the Ru atoms and the 
Ce1 sites, which are centered in the zinc cages.

\section{Conclusion}
\label{concl}

The intermediate-valent compound $ {\rm Ce_2RuZn_4} $ was investigated 
using specific heat and susceptibility measurements as well as first 
principles electronic structure calculations. This material is 
characterized by the presence of two different cerium sites with 
trivalent/intermediate valence ordering. According to the calculations, 
these differences go along with a striking variance in Ce-Ru bonding, 
which is strong at the intermediate-valent cerium sites and rather 
weak at the Ce1 positions at the center of the characteristic zinc 
cages. As a consequence, the LDA+U calculations reveal almost perfectly 
localized magnetic moments of $ {\rm 1.0 \mu_B} $ at the latter sites, 
whereas the Ru-bonded cerium displays no spin polarization. The 
theoretical findings are in full agreement with the experimental data, 
which point to an antiferromagnetic transition at a N\'eel-temperature 
of $ T_{\rm N} = 2 $\,K.

\section{Acknowledgements}

This work was supported by the Deutsche Forschungsgemeinschaft through 
SFB 484 and SCHE 487/7-1. 
WH is indebted to the Fonds der Chemischen Industrie for a PhD stipend.

%
%


\begin{thebibliography}{99}

\bibitem{emsley99}
J.\ Emsley, 
{\em The Elements} (Oxford University Press, Oxford 1999). 

\bibitem{compton59}
V.\ B.\ Compton and B.\ T.\ Matthias, 
Acta Crystallogr.\ {\bf 12}, 651 (1959). 

\bibitem{wilhelm70}
M.\ Wilhelm and B.\ Hillenbrand, 
J.\ Phys.\ Chem.\ Solids {\bf 31}, 559 (1970). 

\bibitem{kurenbaeva07}
Z.\ M.\ Kurenbaeva, A.\ I.\ Tursina, E.\ V.\ Murashova, S.\ N.\ Nesterenko, 
A.\ V.\ Gribanov, Y.\ Seropegin, and H.\ No\"el, 
J.\ Alloys Comp.\ {\bf 442}, 86 (2007). 

\bibitem{murashova07}
E.\ V.\ Murashova, Z.\ M.\ Kurenbaeva, A.\ I.\ Tursina, H.\ No\"el, 
P.\ Rogl, A.\ V.\ Grytsiv, A.\ V.\ Gribanov, G.\ Giester, and Y.\ Seropegin,
J.\ Alloys Comp.\ {\bf 442}, 89 (2007). 

\bibitem{tursina07}
A.\ I.\ Tursina, Z.\ M.\ Kurenbaeva, A.\ Gribanov, H.\ No\"el, T.\ Roisnel, 
and Y.\ Seropegin,
J.\ Alloys Comp.\ {\bf 442}, 100 (2007). 

\bibitem{murashova08}
E.\ V.\ Murashova, A.\ I.\ Tursina, Z.\ M.\ Kurenbaeva, A.\ V.\ Gribanov, 
and Y.\ Seropegin,
J.\ Alloys Comp.\ {\bf 454}, 206 (2008). 

\bibitem{riecken07}
J.\ F.\ Riecken, W.\ Hermes, B.\ Chevalier, R.-D.\ Hoffmann, 
F.\ M.\ Schappacher, and R.\ P\"ottgen,
Z.\ Anorg.\ Allg.\ Chem.\ {\bf 633}, 1094 (2007). 

\bibitem{matar07}
S.\ F.\ Matar, J.\ F.\ Riecken, B.\ Chevalier, R.\ P\"ottgen, A.\ F.\ Al Alam,
and V.\ Eyert, 
Phys.\ Rev.\ B.\ {\bf 76} 174434 (2007). 

\bibitem{mishra08}
R.\ Mishra, W.\ Hermes, U.\ Ch.\ Rodewald, R.-D.\ Hoffmann, and R.\ P\"oöttgen,
Z.\ Anorg.\ Allg.\ Chem.\ {\bf 634}, 470 (2008). 

\bibitem{poettgen99}
R.\ P\"ottgen, T.\ Gulden, and A.\ Simon, 
GIT Labor-Fachz.\ {\bf 43}, 133 (1999). 

\bibitem{kussmann98}
D.\ Ku\ss mann, R.-D.\ Hoffmann, and R.\ P\"ottgen 
Z.\ Anorg.\ Allg.\ Chem.\ {\bf 624}, 1727 (1998). 

\bibitem{yvon77}
K.\ Yvon, W.\ Jeitschko, and E.\ Parth\'e, 
J.\ Appl.\ Crystallogr.\ {\bf 10}, 73 (1977). 

\bibitem{bachmann72}
R.\ Bachmann, F.\ J.\ DiSalvo, T.\ H.\ Geballe, R.\ L.\ Greene, 
R.\ E. Howard, C.\ N.\ King, H.\ C.\ Kirsch, K.\ N.\ Lee, 
R.\ E.\ Schwall, H.\ U.\ Thomas, and R.\ B.\ Zubeck, 
Rev.\ Sci.\ Instrumen.\ {\bf 43}, 205 (1972). 

\bibitem{aswrev}
V.\ Eyert, 
Int.\ J.\ Quantum Chem.\, {\bf 77}, 1007 (2000).

\bibitem{aswbook}
V.\ Eyert,
{\em The Augmented Spherical Wave Method -- A Comprehensive Treatment},
Lect.\ Notes Phys.\ {\bf 719} (Springer, Berlin Heidelberg 2007).

\bibitem{sgo}
V.\ Eyert and K.-H.\ H\"ock, 
Phys.\ Rev.\ B {\bf 57}, 12727 (1998).

\bibitem{mixpap}
V.\ Eyert, 
J.\ Comp.\ Phys.\ {\bf 124}, 271 (1996).

\bibitem{bloechl94}
P.\ E.\ Bl\"ochl, O.\ Jepsen, and O.\ K.\ Andersen,
Phys.\ Rev.\ B {\bf 49}, 16223 (1994).

\bibitem{fpasw}
V.\ Eyert, to be published.

\bibitem{msm88}
M.\ S.\ Methfessel, Phys.\ Rev.\ B {\bf 38}, 1537 (1988).





\end{thebibliography}
\end{document}